\documentclass[runningheads]{llncs}
\usepackage{multirow}
\usepackage{makecell}
\usepackage{booktabs}
\usepackage{bbding}
\usepackage[misc]{ifsym}
\usepackage{graphicx}
\usepackage{subfigure}
\usepackage{bm}
\usepackage[numbers]{natbib}
\usepackage{indentfirst}

\usepackage[colorlinks,linkcolor=blue,anchorcolor=blue,citecolor=blue]{hyperref}

\usepackage{xcolor}

\usepackage{multirow}

\begin{document}{}
	\title{\textsc{Oriole}: Thwarting Privacy against Trustworthy Deep Learning Models}
	%
	%
	
	\author{Liuqiao Chen\inst{1}, Hu Wang\inst{2}, Benjamin Zi Hao Zhao\inst{3}, \\ Minhui Xue\inst{2}, \and Haifeng Qian\inst{1(}\Envelope\inst{)}
	}
	\authorrunning{Chen et al.}
	%
	\institute{$^1$ East China Normal University, China\linebreak
		$^2$ The University of Adelaide, Australia\\
		$^3$ The University of New South Wales and Data61-CSIRO, Australia
	}
	\maketitle             
	\begin{abstract}
		Deep Neural Networks have achieved unprecedented success in the field of face recognition such that any individual can crawl the data of others from the Internet without their explicit permission for the purpose of training high-precision face recognition models, creating a serious violation of privacy. Recently, a well-known system named Fawkes~\cite{shan2020fawkes} (published in USENIX Security 2020) claimed this privacy threat can be neutralized by uploading cloaked user images instead of their original images. In this paper, we present \textsc{Oriole}, a system that combines the advantages of data poisoning attacks and evasion attacks, to thwart the protection offered by Fawkes, by training the attacker face recognition model with multi-cloaked images generated by \textsc{Oriole}. Consequently, the face recognition accuracy of the attack model is maintained and the weaknesses of Fawkes are revealed. Experimental results show that our proposed \textsc{Oriole} system is able to effectively interfere with the performance of the Fawkes system to achieve promising attacking results. Our ablation study highlights multiple principal factors that affect the performance of the \textsc{Oriole} system, including the DSSIM perturbation budget, the ratio of leaked clean user images, and the numbers of multi-cloaks for each uncloaked image. We also identify and discuss at length the vulnerabilities of Fawkes. We hope that the new methodology presented in this paper will inform the security community of a need to design more robust privacy-preserving deep learning models.
		
	\end{abstract}
	\section{Introduction}
	Facial Recognition is one of the most important biometrics of mankind and is frequently used in daily human communication~\cite{akbari2010performance}. Facial recognition, as an emerging technology composed of detection, capturing and matching, has been successfully adapted to various fields: photography~\cite{rasti2016convolutional}, video surveillance~\cite{bashbaghi2019deep}, and mobile payments~\cite{vazquez2016face}. With the tremendous success gained by deep learning techniques, current deep neural facial recognition models map an individual's biometric information into a feature space and stores them as faceprints. Consequently, features of a live captured image are extracted for comparison with the stored faceprints. Currently, many prominent vendors offer high-quality facial recognition tools or services, including NEC~\cite{Nec:2020}, Aware~\cite{Aware:2020}, Google~\cite{Google:2020}, and Face++~\cite{Face++:2020} (a Chinese tech giant Megvii). According to an industry research report ``Market Analysis Repo''~\cite{MarketAnalysisReport:2020}, the global facial recognition market was valued around \$3.4 billion in 2019 and is anticipated to expand with a compound annual growth rate (CAGR) of 14.5\% from 2020 to 2027. Along with the universality of facial recognition technology, the concerns of privacy leakage and security breaches continue to grow. According to Kashmir Hill~\cite{ClearviewAI:2020}, a start-up, Clearview AI, scrapes in excess of three billion images from the Internet, off platforms such as Facebook, Instagram and LinkedIn without users' consent, in order to build tools for revealing individual's identity from their images. It is clear that the misuse of the face recognition technology will create great threats against user's privacy. 
	
	Despite the widespread use of facial recognition technology, it is still in its infancy and unresolved issues of security and privacy will worsen in the wake of big data. One act to safeguard user photos from facial recognition model training without consent is proposed by SAND Lab at the University of Chicago. SAND Lab proposed a protection system Fawkes~\cite{shan2020fawkes} (an article published in USENIX Security 2020). The Fawkes system ``cloaks'' a user's original photos to fool the deep learning face recognition models by adding imperceptible perturbations. Fawkes reports remarkable results against state-of-the-art facial recognition services from Microsoft (Azure Face), Amazon (Rekognition), and Face++~\cite{shan2020fawkes}.
	
	In this paper, we present \textsc{Oriole}, a system designed to render the Fawkes system ineffective. In Fawkes, the target class is selected from the public dataset. In contrast, \textsc{Oriole} implements a white-box attack to artificially choose multiple targets and acquire the corresponding multiple cloaked images of leaked user photos. With the help of the proposed multi-cloaks, the protection of Fawkes becomes fragile. To do so, the attacker utilizes the multi-cloaks to train the face recognition model. During the test phase, after the original user images are collected, the attacker inputs the Fawkes cloaked image into the model for face recognition. As a result, in the feature space, the features of cloaked photos will inevitably fall into the range of marked multi-cloaks. Therefore, the user images can still be recognized even if they are cloaked by Fawkes. We also highlight the intrinsic weakness of Fawkes:  The imperceptibility of images before and after cloaking is limited when encountering high-resolution images, as cloaked images may include spots, acne, and even disfigurement. This will result in the reluctance of users to upload their disfigured photos. 
	
	In summary, our main contributions in this paper are as follows:
	
	\begin{itemize}
\item {\bfseries \itshape The Proposal of Oriole.}
	We design, implement, and evaluate \textsc{Oriole}, a neural-based system that makes attack models indifferent to the protection of Fawkes. Specifically, in the training phase, we produce the most relevant multi-cloaks according to the leaked user photos and mix them into the training data to obtain a face recognition model. During the testing phase, when encountering uncloaked images, we first cloak them with Fawkes and then feed them into the attack model. By doing so, the user images can still be recognized even if they are protected by Fawkes.
	
\item {\bfseries \itshape Empirical Results.}
	We provide experimental results to show the effectiveness of \textsc{Oriole} in the interference of Fawkes. We also identify multiple principle factors that affect the performance of the \textsc{Oriole} system, including the DSSIM perturbation budget, the ratio of leaked clean user images, and the number of multi-cloaks for each uncloaked image. Furthermore, we identify and discuss at length the intrinsic vulnerability of Fawkes to deal with high-resolution images.
	
	\end{itemize}
	
	\section{Related Work}
\begin{figure}[t]
		\includegraphics[width=0.85\textwidth]{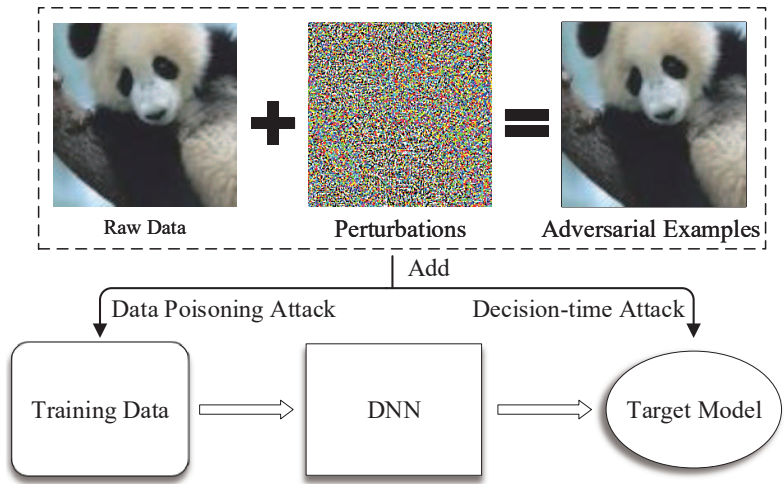}
		\centering
		\caption{The differences between data poisoning attacks and decision-time attacks. Data poisoning attacks modify the training data before the model training process. In contrast, Decision-time attacks are performed after model training to induce the model make erroneous predictions.
		} \label{DifferenceBetweenDataPoisoningAndDecisionTime}
	\end{figure}

	In this section, we briefly introduce defense strategies against data poisoning attacks and decision-time attacks.
	Figure~\ref{DifferenceBetweenDataPoisoningAndDecisionTime} highlights the differences between data poisoning attacks and decision-time attacks. We then introduce the white-box attacks. The Fawkes system is detailed at the end of this section.
	
	\subsection{Defending against Data Poisoning Attacks} 
	\noindent
	In the scenario of data poisoning attacks, the model’s decision boundary will be shifted due to the injection of adversarial data points into training set. The intuition behind it is that the adversary deliberately manipulates the training data since the added poisoned data has vastly different distribution with the original training data~\cite{wen2021great,li2021hidden,9186317,li2020deep,wen2020palor,xu2021explainability}. Prior research primarily involves two common defense strategies. First, anomaly detection models~\cite{wang2019unsupervised} function efficiently if the injected data has obvious differences compared to the original training data. Unfortunately, anomaly detection models become ineffective if the adversarial examples are inconspicuous. Similar ideas have been utilized in digital watermarking or data hiding~\cite{zhang2020robust}. Second, it is common to analyze the impact of newly added training samples according to the accuracy of models. For example, Reject On Negative Impact (RONI) was proposed against spam filter poisoning attacks, while Target-aware RONI (tRONI) builds on the observation of RONI failing to mitigate targeted attacks~\cite{suciu2018does}. 
	Other notable methods include TRIM~\cite{jagielski2018manipulating}, STRIP~\cite{gao2019strip}, and more simply, human analysis on training data likely to be attacked~\cite{DBLP:conf/aaai/MeiZ15}.
	
	\subsection{Defending against Decision-time Attacks}
	\noindent
	In decision-time attacks, assuming that the model has already been learned, the attacker leads the model to produce erroneous predictions by making reactive changes to the input. Decision-time attacks can be divided into several categories. Within these attacks, the most common one is the evasion attack. 
	
	We shall present the most conventional evasion attack, which can be further broken down into five categories: Gradient-based attacks~\cite{carlini2017adversarial,DBLP:conf/aaai/ChenSZYH18,madry2017towards}, Confidence score attacks~\cite{ilyas2018black,chen2017zoo}, Hard label attacks~\cite{brendel2017decision}, Surrogate model attacks~\cite{zugner2018adversarial} and Brute-force  attacks~\cite{engstrom2019exploring,hendrycks2018benchmarking,DBLP:journals/corr/abs-1901-10513}). Undoubtedly, adversarial training is presently one of the most effective defenses. Adversarial samples, correctly labeled, are added to the training set to enhance model robustness. Input modification~\cite{liao2018defense}, extra classes~\cite{hosseini2017blocking} and detection~\cite{meng2017magnet,grosse2017statistical} are common defense techniques against evasion attacks. Alternative defenses against decision-time attacks involve iterative retraining~\cite{li2018evasion,tong2019improving}, and decision randomization~\cite{shah2019evaluating}.
	
	\subsection{White-box Attacks}
	The adversary has full access to the target DNN model's parameters and architecture in white-box attacks. For any specified input, the attacker can calculate the intermediate computations of each step as well as the corresponding output. Therefore, the attacker can leverage the outputs and the intermediate result of the hidden layers of the target model to implement a successful attack. Goodfellow et al.~\cite{goodfellow2014explaining} introduce a fast gradient sign method (FGSM) to attack neural network models with perturbed adversarial examples according to the gradients of the loss with respect to the input image. The adversarial attack proposed by Carlini and Wagner is by far one of the most efficient white-box attacks~\cite{carlini2017towards}.
	
	\subsection{Fawkes} 
	Fawkes~\cite{shan2020fawkes}, provides privacy protections against unauthorized training of models by modifying user images collected without consent by the attacker. Fawkes achieves this by providing as simple means for users to add imperceptible perturbations onto the original photos before uploading them to social media or public web. When processed by Fawkes, the features representing the cloaked and uncloaked images are hugely different in the feature space but are perceptually similar. The Fawkes system cloaks images by choosing (in advance) a specific target class that has a vast difference to the original image. Then it cloaks the clean images to obtain the cloaked images with great alterations to images' feature representations, but indistinguishable for naked eyes. When trained with these cloaked images, the attacker's model would produce incorrect outputs when encountering clean images. However, Fawkes may be at risk of white-box attacks. If the adversary can obtain full knowledge of the target model’s parameters and architecture, for any specified input, the attacker can calculate any intermediate computation and the corresponding output. Thus, the attackers can leverage the results of each step to implement a successful attack.
	
	\section{Design Overview}
For a clean image $x$ of a user Alice, \textsc{Oriole} produces multi-cloaks by adding pixel-level perturbation to $x$ when choosing multiple targets dissimilar to Alice in the feature space. That is, we first need to determine the target classes and their numbers for each user; then, we shall generate multi-cloaks with these selected classes. The process is detailed in Section~\ref{modeltraining}.
	
\begin{figure}[t]
		\includegraphics[width=1.0\textwidth]{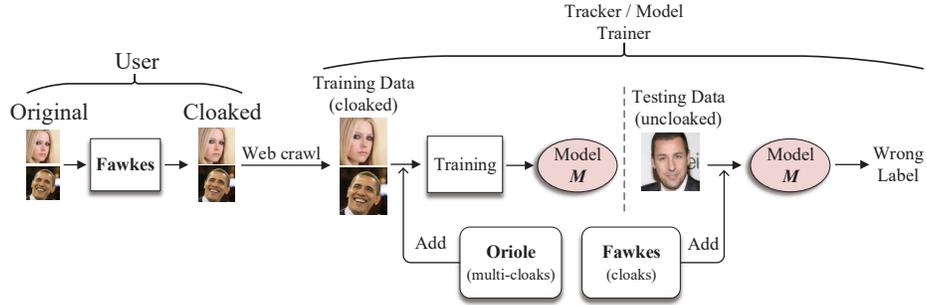}
		\centering
		\caption{The proposed \textsc{Oriole} system is able to successfully recognize faces, even with the protection of Fawkes. \textsc{Oriole} achieves this by combining the concepts of data poisoning attacks and evasion attacks.} \label{Fawkes}
	\end{figure}

	Figure~\ref{Fawkes} illustrates the overview of the proposed \textsc{Oriole} system, together with both its connection and the differences with Fawkes. In the proposed \textsc{Oriole}, the implementation is divided into two stages: training and testing. In the training phase, the attacker inserts the multi-cloaks generated by the \textsc{Oriole} system into their training set. After model training, upon encountering clean user images, we use Fawkes to generate cloaked images; the cloaked images are then fed into the trained face recognition model to complete the recognition process.  \textsc{Oriole} has significant differences with Fawkes. On one hand, we adopt a data poisoning attack scheme against the face recognition model by modifying images with generated multi-cloaks. On the other hand, an evasion attack (to evade the protection) is applied during testing by converting clean images to their cloaked version before feeding them into the unauthorized face recognition model. Although the trained face recognition model cannot identify users in clean images, it can correctly recognize the cloaked images generated by Fawkes and then map them back to their ``true'' labels.
	
	\section{The \textsc{Oriole} System Against Fawkes}
\begin{figure}[th]
		\includegraphics[width=1.0\textwidth]{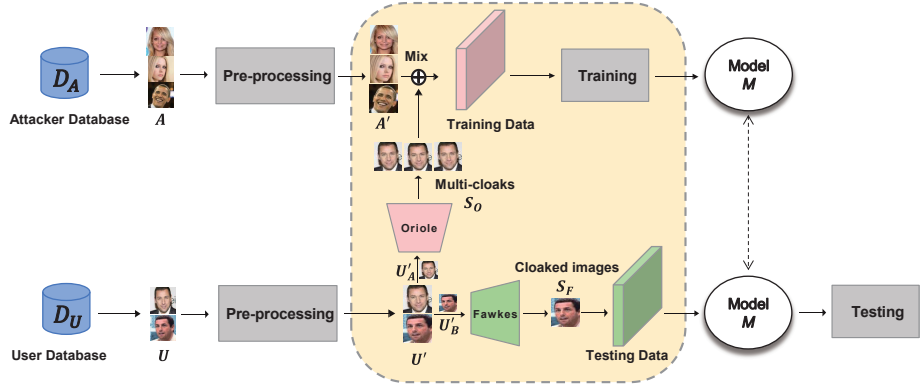}
		\centering
		\caption{
			The overall process of the proposed \textsc{Oriole}. The process includes both the training and testing stages. Images $U$ taken from the leaked user database $D_U$ are divided into two parts ($U_A^{'}$ and $U_B^{'}$) after preprocessing. In the training phase, the attacker $A$ mixes the generated multi-cloaks $S_O$ into training data. After training, the face recognition model $M$ is obtained. During the testing phase, the attacker $A$ first converts the clean images $U_B^{'}$ into cloaked images $S_F$ and then pipes them into the trained model $M$ to obtain a correct prediction.
		} \label{OrioleSystemOverview}
	\end{figure}

	We now elaborate the design details of \textsc{Oriole}. We refer to the illustration of the \textsc{Oriole} process in Figure~\ref{OrioleSystemOverview}. Recall that the application of \textsc{Oriole} is divided into a training phase and a testing phase. The training phase can be further broken down into two steps. In the first step, the attacker $A$ launches a data poisoning attack to mix the multi-cloaks into the training data (recall that the training data is collected without consent and has been protected by Fawkes). Then, the unauthorized facial recognition model $M$ is trained on the mixed training data of the second step. At test time, as evasion attacks, the attacker $A$ first converts the clean testing images to the cloaked version by applying Fawkes and the cloaked version is presented to the trained model $M$ for identification. From Figure~\ref{OrioleSystemOverview}, images making up the attacker database $D_A$ can be downloaded from the Internet as training data, while the user database $D_U$ provides the user $U$ with leaked and testing data. After obtaining the input images from the database, we adopt MTCNN~\cite{zhang2016joint} for accurate face detection and localization as the preprocessing module~\cite{zhang2016joint,xiang2017joint}. It outputs standardized images that only contain human faces with a fixed size. At the training phase, the attacker $A$ mixes the processed images of $A^{'}$ and multi-cloaks $S_O$ of the user $U$ into training set to train the face recognition model $M$. At the testing phase, the attacker $A$ first converts the preprocessed clean images $U_{B}^{'}$ into the cloaked images $S_F$, followed by the same procedure as described in Fawkes; then, the attacker $A$ pipes $S_F$ into the trained model $M$ to fetch the results. 
	
	\subsection{Model Training}
	\label{modeltraining}
	
	We assume that a user $U$ has converted his/her clean images $U_B$ into their cloaked form for privacy protection. However, the attacker $A$ has collected some leaked clean images of the user $U$ in advance, denoted as $U_A$. As shown in Figure~\ref{OrioleSystemOverview}, this leaked user dataset $U$ consists of data needed $U_A$ and $U_B$.
	In the proposed \textsc{Oriole} system, $U_A$ is utilized for obtaining multi-cloaks $S_O$, which contains a target set $T_M$ with $m$ categories out of $N$ categories.\footnote{\url{http://mirror.cs.uchicago.edu/fawkes/files/target\_data/}} Here, we denote $G(X,m)$ as the new set composed of the target classes corresponding to the first $m$ largest element values in set $X$, where $X$ contains the minimum distance between the feature vector of users and the centroids of $N$ categories (see Eq.~\ref{T_M}). The $L_2$ distances are measured between the image feature in the projected space $\Phi(\cdot)$ to the centroids of $N$ categories, and then the top $m$ targets are selected.

	\begin{equation}
		X = \bigcup_{k=1}^{N} \{d\mid d = \min_{x\in U_B} \left(Dist(\Phi(x),C_k)\right)\},
	\end{equation}
	
	\begin{equation}
	    \label{T_M}
		T_M=G\left(X,m\right)=\{T_1,T_2,\cdots,T_m\}=\bigcup_{i=1}^{m}T_i,
	\end{equation}
	\noindent
	where $C_k$ represents the centroid of a certain target and $\Phi$ is the feature projector~\cite{shan2020fawkes}. Besides, the distance calculation function adopts $L_2$ distance. Next, the calculation of a cloak $\delta\left(x,x_{T_i}\right)$ is defined as:
	
	\begin{equation}
		\delta(x,X_{T_i})=\min_\delta Dist(\Phi\left(x_{T_i}\right),\Phi (x \oplus \delta\left(x,x_{T_i}\right))),
	\end{equation}
	
	\noindent
	where $\delta$ subjects to $\left|\delta(x,x_{T_i})\right|<\rho$, and  $\left|\delta(x,x_{T_i})\right|$ is calculated by DSSIM (Structure Dis-Similarity Index)~\cite{wang2018great,wang2003multiscale} and $\rho$ is the perturbation budget. Then we can obtain the multi-cloaks $S_O$ as follows:
	
	\begin{equation}
		S_O = \bigcup_{i=1}^{m}\{s\mid s=x\oplus\delta(x,x_{T_i})\}\label{S_O},
	\end{equation}
	where multi-value $m$ is a tunable hyper-parameter. $m$ decides the number of multi-cloaks produced for each clean image.
	
	Instead of training the model $M$ with clean data, the attacker $A$ mixes the multi-cloaks $S_O$ calculated from Equation~\ref{S_O} with the preprocessed images $U^{'}_{A}$ to form the training set. The deep convolutional face recognition model $M$ is trained~\cite{schroff2015facenet}.
	
	\subsection{Model Testing}
	The last stage of \textsc{Oriole} is model testing. Unlike Fawkes, we do not directly apply clean images to the attack model. Instead, \textsc{Oriole} first makes subtle changes to the clean images before faces identification inference. Specifically, we implement the subtle changes through cloaking images from processed user images $U_{B}^{'}$. Conceptually, the feature vectors of cloaked images $S_F$ will fall into the marked feature space of multi-cloaks $S_O$. Then, the trained model $M$ is able to correctly identify users through cloaked images $S_F$.
	
\begin{figure}[t]
		\centering
		\includegraphics[width=1.0\textwidth]{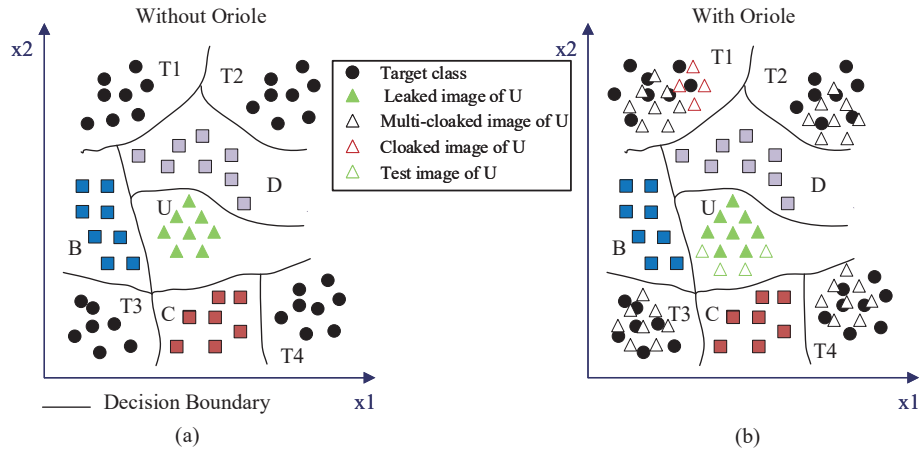}
		\caption{The intuition behind why \textsc{Oriole} can help the attacker $A$ successfully identify the user $U$ even with the protection of Fawkes. We denote the process on a simplified 2D feature space with seven user classes $B, C, D, T_1, T_2, T_3, T_4$ and $U$. 
			Figures~(a) and (b) represent the decision boundaries of the model trained on $U$'s clean photos and multi-cloaks respectively (with four targets). The white triangles represent the multi-cloaked images of $U$ and the red triangles are the cloaked images of $U$. \textsc{Oriole} works as long as cloaked testing images fall into the same feature space of the multi-cloaked leaked images of $U$.
		}\label{Decision_Boundary}
	\end{figure}

	Figure~\ref{Decision_Boundary} illustrates the intuition behind the \textsc{Oriole} system. For the purposes of demonstration, we assume the number of multi-value $m$ equals to four. To put differently, we shall assume that Fawkes will select one of four targets for cloaking, from which the proposed \textsc{Oriole} system will attempt to obtain multi-cloaks associated with all four targets with a small number of the user $U$'s leaked photos. In this scenario, we successfully link the four feature spaces of our four target classes ($T_1,T_2,T_3$ and $T_4$) with the user $U$. Thus, when it comes to a new and clean image of $U$, we first cloak it with Fawkes. The cloaked version user images will inevitably fall into one of the marked feature spaces of the multi-cloaks ($T_1$ has been chosen for illustration in Figure~\ref{Decision_Boundary}(b). See the hollow green and red triangles for the clean and cloaked image features, respectively). As the cloaked image features lie in $T_1$, and the multi-cloak trained model now associates $T_1$ (and $T_2,T_3,T_4$) as $U$, the attacker can correctly identify a user's identity even with the protection of Fawkes.
	
	We finally discuss the performance of \textsc{Oriole} when target classes are included and not included in the training data, respectively. We further observe that, no matter whether the number of target classes $m$ is included in the training set or not, the \textsc{Oriole} system still functions effectively to thwart protections offered by Fawkes. In Figure~\ref{Decision_Boundary}, assuming that the feature vectors of the cloaked testing image are located in the high dimensional feature space of $T_1$. 
	We first consider when target users of $T_1$ are not included in the attack model training process. We are able to map the user $U$ to the feature space of $T_1$ through the leaked images of the user $U$ that were used to generate multi-cloaks. 
	Furthermore, \textsc{Oriole} still works when images of the target class $T_1$ are included in the training set. Even if the cloaked images of $U$ are detected as $T_1$, but the setting of Fawkes ensures that the cloaks of $T_1$ occupy another area within the feature space that will not overlap with $T_1$. Thus, this special case will not interfere the effectiveness of \textsc{Oriole}.
	
\section{Experiments}
\subsection{Datasets and Models}
\noindent
We implemented our \textsc{Oriole} system on three popular image datasets against the Fawkes system. In our implementation, considering the size of the three datasets, we took the smallest PubFig83~\cite{pinto2011scaling} as the user dataset, while the larger VGGFace2~\cite{cao2018vggface2} and CASIA-WebFace~\cite{yi2014learning} were prepared for the attacker to train two face recognition models. In addition, we artificially created a high-definition face dataset to benchmark the data constraints surrounding the imperceptibility of the Fawkes system.\footnote{Our source code is publicly available at \url{https://git.io/JsWq7}.} 

\paragraph{\textbf{PubFig83}~\cite{pinto2011scaling}.}
PubFig83 is a well-known dataset for face recognition research. It contains 13,838 cropped facial images belonging to 83 celebrities, each of which has at least 100 pictures. In our experiment, we treat PubFig83 as a database for user sample selection, due to its relative small number of tags and consistent picture resolution.

\paragraph{\textbf{CASIA-WebFace}~\cite{yi2014learning}.}
CASIA-WebFace dataset is the largest known public dataset for face recognition, consisting a total of 903,304 images in 38,423 categories.

\paragraph{\textbf{VGGFace2}~\cite{cao2018vggface2}.}
VGGFace2 is a large-scale dataset containing 3.31 million images from 9131 subjects, with an average of 362.6 images for each subject. All images on VGGFace2 were collected from the Google Image Search and distributed as evenly as possible on gender, occupation, race, etc.

\paragraph{\textbf{Models:} \bm{$M_V$} and \bm{$M_{CW}$}.}
We chose VGGFace2 and CASIA to train face recognition models separately for real-world attacker simulation. In the preprocessing stage, MTCNN~\cite{zhang2016joint} is adopted for face alignment and Inception-ResNet-V1~\cite{szegedy2017inception} selected as our model architecture, and we then completed the model training process on a Tesla P100 GPU, with Tensorflow r1.7. An Adam optimizer with a learning rate of -1 is used to train models over 500 epochs. Here, we denote the models trained on the VGGFace2 and CASIA-WebFace datasets as $M_V$ and $M_{CW}$, the LFW accuracy of these models achieved $99.05\%$ and $99.65\%$, respectively.

\subsection{Experimental Evaluation}\label{expeval}
	Similar to the Fawkes system, the proposed \textsc{Oriole} system is designed for a user-attacker scenario, whereby the attacker trains a powerful model through a huge number of images collected on the Internet. The key difference is that \textsc{Oriole} assumes the attacker $A$ is able to obtain a small percentage of leaked clean images of user $U$. Through the evaluation of the \textsc{Oriole} system, we discover the relevant variables affecting the attack capability of the \textsc{Oriole} system. In this case, we define a formula for facial recognition accuracy evaluation in Equation~\ref{EqAcc}, where $R$ represents the ratio of the user's multi-cloaks in the training data. The ranges of $R$ and $\rho$ are both set to $[0,1]$, and the parameter $m$ (number of multi-cloaks) is subject to the inequality: $0<m\ll{N}$, where $N=18,947$ is the total number of target classes in the public dataset.
	
	\begin{equation}
		Accuracy = k\frac{R\cdot m}{\rho}\label{EqAcc}
	\end{equation}
	
	Throughout our experimental evaluation, the ratio between the training data and testing data is fixed at 1:1 (see Section~\ref{expeval} for the motivation behind this ratio).
	
	\paragraph{\textbf{Comparison between Fawkes and \textsc{Oriole}.}}
	We start by reproducing the Fawkes system against unauthorized face recognition models. Next, we employed the proposed \textsc{Oriole} scheme to invalidate the Fawkes system. We shall emphasize that the leaked data obtained associated with the user will not be directly used for training the attack model. Instead, we insert multi-cloaks actively produced by \textsc{Oriole} into the training process, which presents a significant difference in the way adversary training schemes deal with leaked data.

\begin{figure}[t]
		\centering
		\includegraphics[width=1\textwidth]{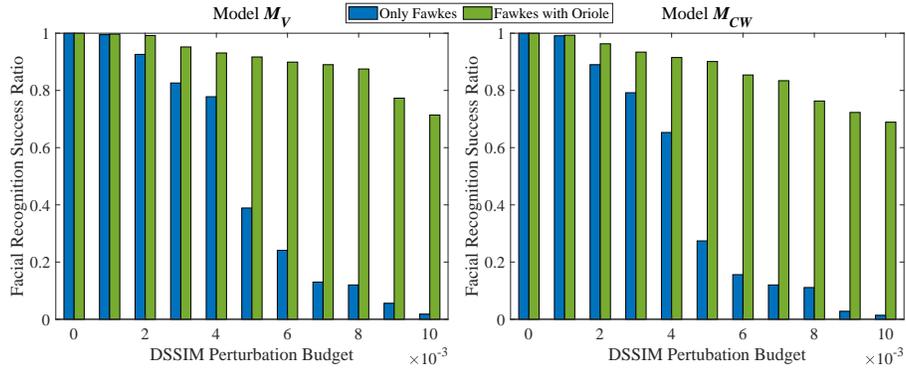}
		\caption 
		{Evaluation of the impact on \textsc{Oriole} against Fawkes through two models $M_V$ and $M_{CW}$. The two figures depict the performance of the face recognition model $M$ with Fawkes and equipped with \textsc{Oriole}. There are clear observations from the two figures: the larger the DSSIM perturbation budget $\rho$, the higher the resulting face recognition accuracy obtained from model $M$. Additionally, it demonstrates that our proposed \textsc{Oriole} system can successfully bypass protections offered by Fawkes.
		}\label{FawkesAndOriole}
	\end{figure}

	In particular, we randomly select a user $U$ with 100 images from PubFig83 and divided their images equally into two non-intersecting parts: $U_A$ and $U_B$, each of which contains 50 images, respectively. We shall evaluate both Fawkes and \textsc{Oriole} in two settings for comparison. In the first setting, we mix the multi-cloaks of the processed $U_A^{'}$ into the training data to train the face recognition model $M$ and test the accuracy of this model $M$ with the processed $U_B^{'}$ in the testing phase (see Figure~\ref{OrioleSystemOverview}). In the second setting, we replace the clean images of $U_A$ with the corresponding cloaked images (by applying Fawkes) to obtain a secondary measure of accuracy. Figure~\ref{FawkesAndOriole} shows the variation in facial recognition accuracy with certain DSSIM perturbation budget, and displays the performance of \textsc{Oriole} against Fawkes protection. We implement this process on two different models: $M_{V}$ and $M_{CW}$. The former training data consists of the leaked images $U_A$ and all images in VGGFace2, while the latter contains the leaked images $U_A$ and all images in CASIA-WebFace. All experiments were repeated three times and the results presented are averages.

    \begin{figure}[t]
		\centering
		\includegraphics[width=0.95\textwidth]{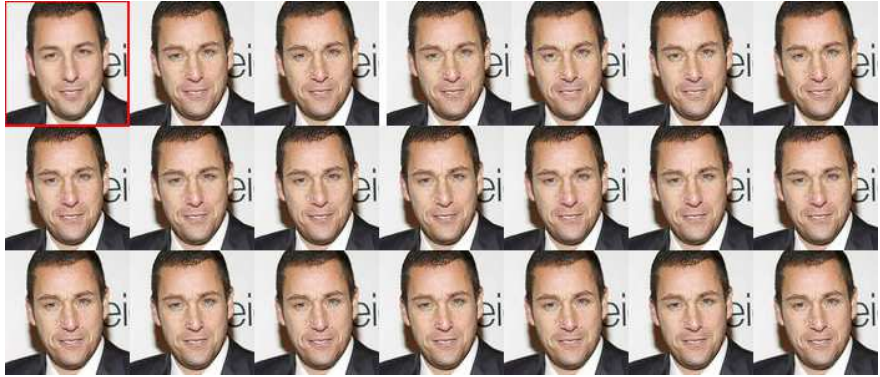}
		\caption{An example of a clean image of the user $U$
		and 20 multi-cloaks produced by \textsc{Oriole}. The uncloaked image has been framed by a red outline.}\label{CleanAndMulti-cloaks}	
	\end{figure}
	
\begin{figure}[t]
		\centering
		\includegraphics[width=1.0\textwidth]{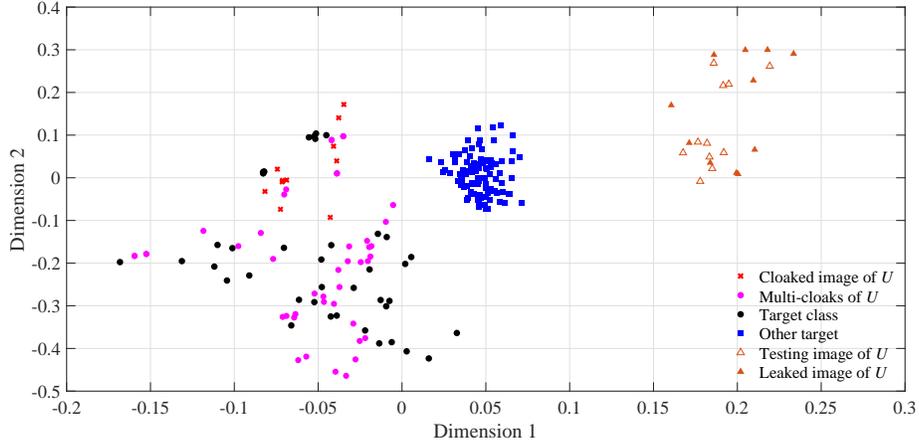}
		\caption{2-Dimensional PCA visualization in our proposed \textsc{Oriole} system. Triangles are user's leaked images (solid) and testing data (hollow), dots are multi-cloaks of leaked images, dots represent multi-cloaks (magenta) and images from target classes (black), red crosses are cloaked images of testing data, blue square are images from another class.}\label{pca}	
	\end{figure}

	It can been seen from Figure~\ref{FawkesAndOriole} that there is a clear trend that the facial recognition ratio of the two models rises significantly as the DSSIM perturbation budget $\rho$ increases from 0.1 to 1. Specifically, \textsc{Oriole} improves the accuracy of the face recognition model $M_V$ from 12.0\% to 87.5\%, while the accuracy of the model $M_{CW}$ increases from 0.111 to 0.763 when parameter $\rho$ is set to 0.008. We notice that the accuracy of the two models $M_V$ and $M_{CW}$ has been improved nearly 7 fold, when compared to the scenario where Fawkes is used to protect privacy. From these results, we empirically find that \textsc{Oriole} can neutralize the protections offered by Fawkes, invalidating its protection of images in unauthorized deep learning models. Figure~\ref{CleanAndMulti-cloaks} shows an uncloaked image and its related multi-cloaks ($\rho=0.008, m=20$). The feature representation of the clean image framed by a red outline is dissimilar from that of the remaining 20 images. Figure~\ref{pca} shows the two-dimensional Principal Component Analysis (PCA) of the face recognition system validating our theoretical analysis (for $\rho=0.008, m=4$). The feature representation of the clean images are mapped to the feature space of the four target classes images through multi-cloaks. We then mark the corresponding feature spaces as part of identity $U$ and identify the test images of $U$ by cloaking them.

	We show the general effectiveness of the proposed \textsc{Oriole} system in Table~\ref{Universality}. We build four models with two different architectures, named Inception-ResNet-V1~\cite{szegedy2017inception} and DenseNet-121~\cite{huang2017densely}, on the two aforementioned datasets. The model, equipped with \textsc{Oriole}, significantly outperforms the model without it across different setups. The experimental results demonstrate that the \textsc{Oriole} system can retain the test accuracy at a higher level of more than 70\% accuracy across all listed settings, even with the protection of Fawkes. For instance, on the CASIA-WebFace dataset with DenseNet-121 as the backbone architecture, \textsc{Oriole} increases the attack success rate from 12.0\% to 87.5\%, significantly boosting the attack effectiveness. 
	
		\begin{table}[t]
		\centering
		\caption{The four models  used in our verification and their classification accuracy on PubFig83. The ``Basic'' column represents the conventional face recognition. The ``Fawkes'' column represents that only Fawkes is used to fool the face recognition model for privacy protection. The  \textsc{Oriole} column represents the performance of \textsc{Oriole}.} \label{Universality}
			\begin{tabular}{c|c|c|c|c}
				\hline
				\multirow{2}*{~~~\textbf{Dataset}~~~} & \multirow{2}*{~~\textbf{Model Architecture}~~} &
				\multicolumn{3}{c}{~~\textbf{Test Accuracy}~~}\\\cline{3-5}
				& & ~~\textbf{Basic}~~ & ~~\textbf{Fawkes}~~ & ~~\textbf{\textsc{Oriole}}~~\\
				\Xhline{3\arrayrulewidth}
				~CASIA-WebFace~ & Inception-ResNet-V1 & 0.973 & 0.111 & 0.763  \\\hline
				~CASIA-WebFace~ & DenseNet-121  & 0.982 & 0.214 & 0.753 \\\hline
				VGGFace2 & Inception-ResNet-V1 & 0.976 & 0.120 & 0.875 \\\hline
				VGGFace2 & DenseNet-121 & 0.964 & 0.117 & 0.714 \\\hline
		\end{tabular}
	\end{table}
	
\begin{figure}[t]
		\centering
		\includegraphics[width=1.0\textwidth]{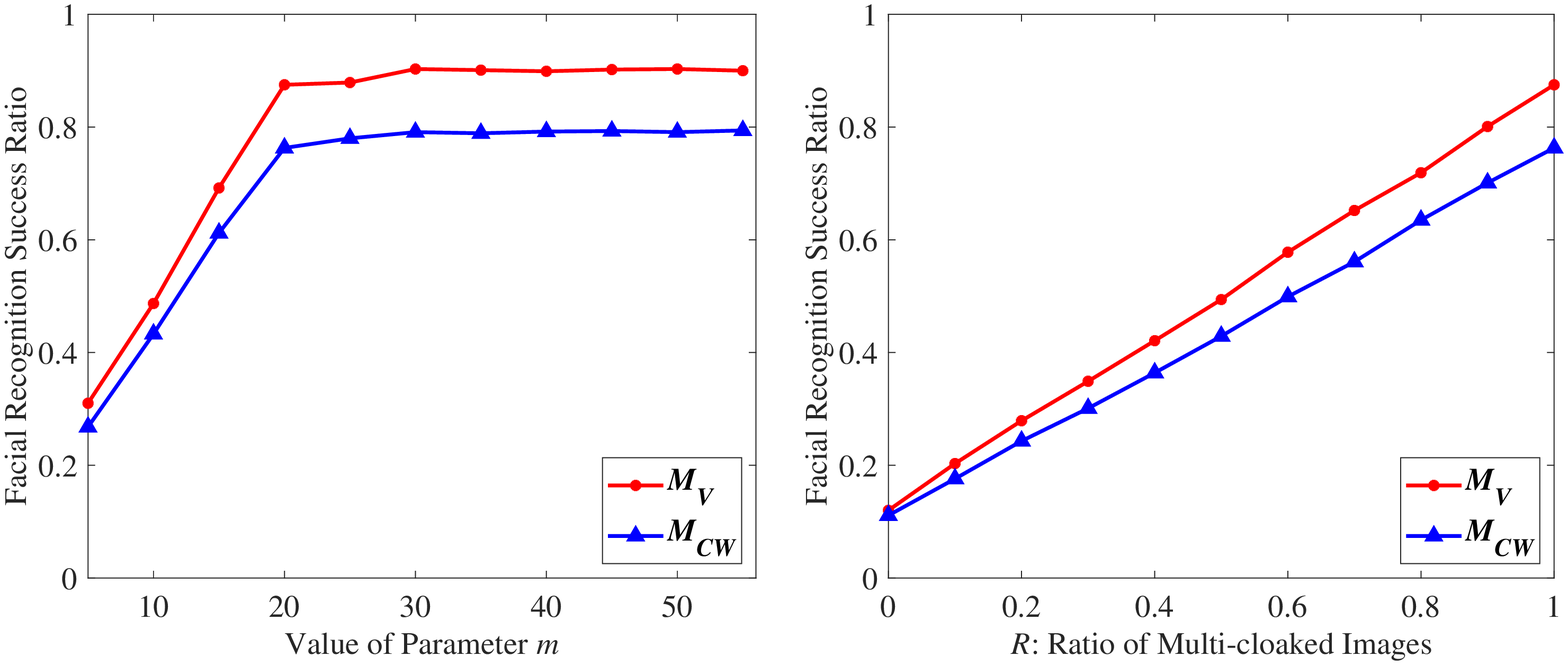}
		\caption{The facial recognition accuracy changes with different ratios of leaked clean images $R$ and numbers of multi-cloaks for each uncloaked image $m$.
		}
		\label{FacialRecognitionRatioOfRAndMulti}
		
	\end{figure}

	\paragraph{\textbf{Main factors contributing to the performance of \textsc{Oriole}.}}
	There are three main factors influencing the performance of \textsc{Oriole}: 1) the DSSIM perturbation budget $\rho$, 2) the ratio of leaked clean images $R$, and 3) the number of multi-cloaks for each uncloaked image $m$. 
	Different DSSIM perturbation budgets $\rho$ have already been discussed in the previous paragraph. We now explore the impact of $R$ and $m$ values on model's performance. Up until this point we have performed experiments with default values of $R$, $m$ and $\rho$ as 1, 20 and 0.008 respectively to enable a fair comparison. 
	From Figure~\ref{FacialRecognitionRatioOfRAndMulti} we can observe the main factors affecting the \textsc{Oriole} system's performance. 
	We observe that the facial recognition success ratio increases monotonically as the number of multi-cloaks $m$ increases, and this rise occurs until $m$ reaches 20, whereby the success ratio plateaus. We can conclude that the facial recognition success ratio grows with the ratio of leaked clean images~$R$. The ratio increases at least three times when $R$ increases from 0.1 to 1.

	\paragraph{\textbf{Model validation.}}
    In order to ensure the validity of \textsc{Oriole}, as a comparative experiment, we respectively evaluate the model $M_V$ and $M_{CW}$ on PubFig83. We divide PubFig83 into 10 training-testing set pairs with different proportions and build classifiers with the help of two pre-trained models. We obtained 20 experimental results 
	depending on which model $M_V$ or $M_{CW}$ was used with ratios selected between 0.1 to 1 shown in Table~\ref{AccuracyOfTwoModels}. The experimental results show that the accuracy of model $M_{V}$ and $M_{CW}$ based on FaceNet increases monotonically as the ratio of the training set to the testing set increases. We can see that both models exceed a $96\%$ recognition accuracy on PubFig83 when the selected the ratio between training and testing sets are 0.5. Consequently, models $M_V$ and $M_{CW}$ are capable of verifying the performance of \textsc{Oriole}.

	\begin{table}[t]
		\centering
		\caption{The test accuracy of models $M_{V}$ (trained on VGGFace2) and $M_{CW}$ (trained on CASIA-WebFace) across different rates of PubFig83. The rate in the first column represents the ratio of the size of training and test sets. The test accuracy is the overall correct classification score for clean images.} \label{AccuracyOfTwoModels}
			\begin{tabular}{c|c|c}
				\hline
				~~\textbf{Rate} ~~& ~~\textbf{Test Accuracy of $M_{V}$ } ~~& ~~\textbf{Test Accuracy of $M_{CW}$}~~ \\ 
				\Xhline{3\arrayrulewidth}
				0.1 & 0.952 & 0.923  \\\hline
				0.2 & 0.963 & 0.947 \\\hline
				0.3 & 0.966 & 0.953 \\\hline
				0.4 & 0.968 & 0.957 \\\hline
				0.5 & 0.969 & 0.961 \\\hline
				0.6 & 0.970 & 0.965 \\\hline
				0.7 & 0.972 & 0.969 \\\hline
				0.8 & 0.976 & 0.973 \\\hline
				0.9 & 0.992 & 0.973 \\
				\hline 
		\end{tabular}
	\end{table}

\section{Discussion}
\subsection{Restricted Imperceptibility of Fawkes}
	Shan et al.~\cite{shan2020fawkes} claim that the cloaked images with small perturbations added are indistinguishable to the naked human eye. However, we show that the imperceptibility of Fawkes is limited due to its inherent imperfection, which is vulnerable to white-box attacks. 
	For practical applications, users tend to upload clear and high-resolution pictures for the purpose of better sharing their life experiences. 
	Through our empirical study, we find that Fawkes is able to make imperceptible changes for low-resolution images, such as the PubFig83 dataset. However, when it comes to high-resolution images, the perturbation between cloaked photos and their originals is plainly apparent. 
	
	To demonstrate the limitations in Fawkes for high-resolution images, we manually collect 54 high-quality pictures covering different genders, ages and regions, whose resolution is more than 300 times (width $\times$ height is larger than 3,000,000 pixels at least) of PubFig83 images. We further conduct an experiment to set the value of perturbation budget $\rho$ to 0.007 and run the optimization process for 1,000 iterations with a learning rate of 0.5, in the same experimental setting as described in Fawkes~\cite{shan2020fawkes}.
	
\begin{figure}[htbp]
		\centering
		\subfigure[uncloaked]{
			\begin{minipage}[b]{0.45\linewidth}
				\centering
				\includegraphics[width=1.0\textwidth]{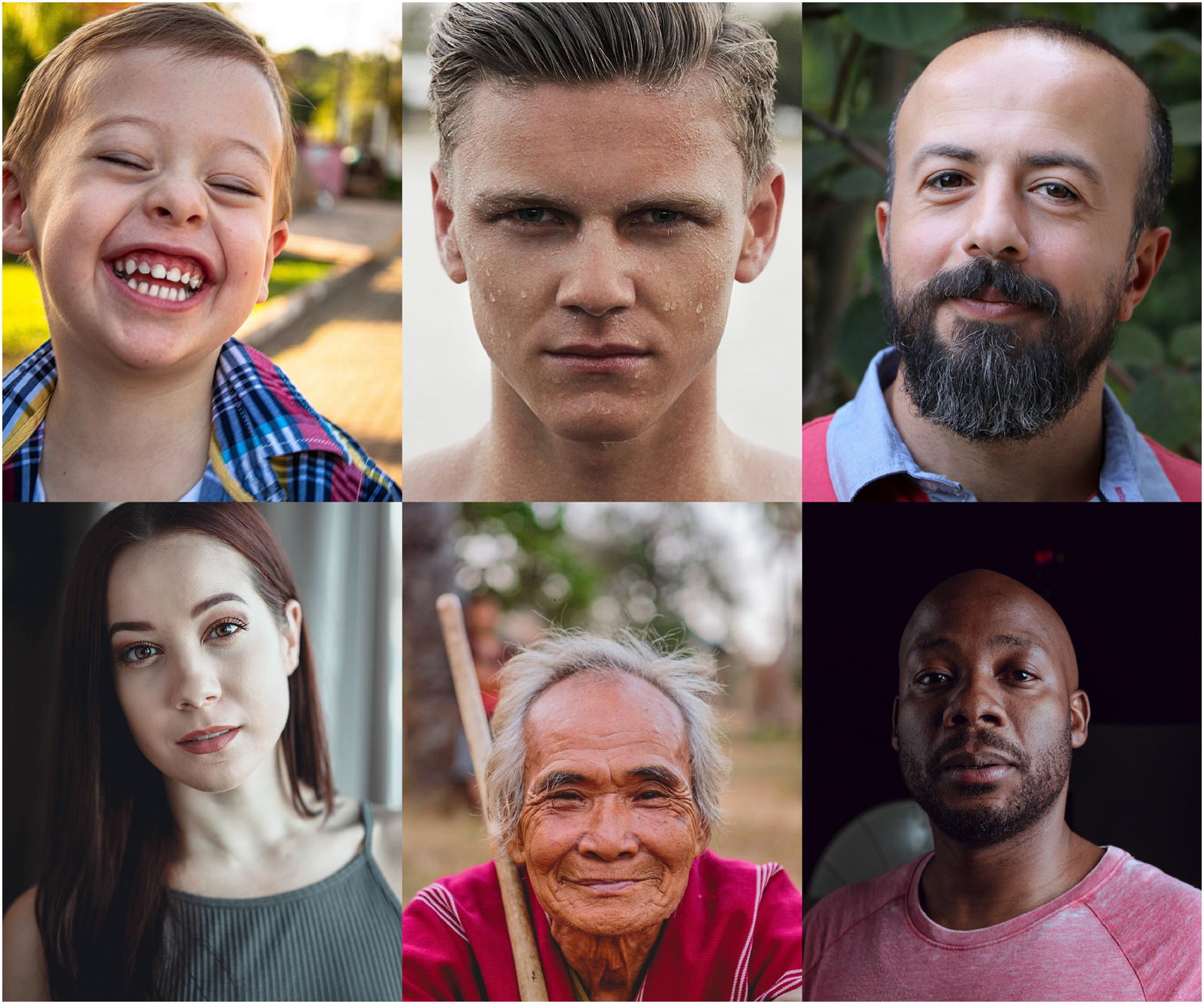}
			\end{minipage}
		}
		\subfigure[cloaked]{
			\begin{minipage}[b]{0.45\linewidth}
				\centering
				\includegraphics[width=1.0\textwidth]{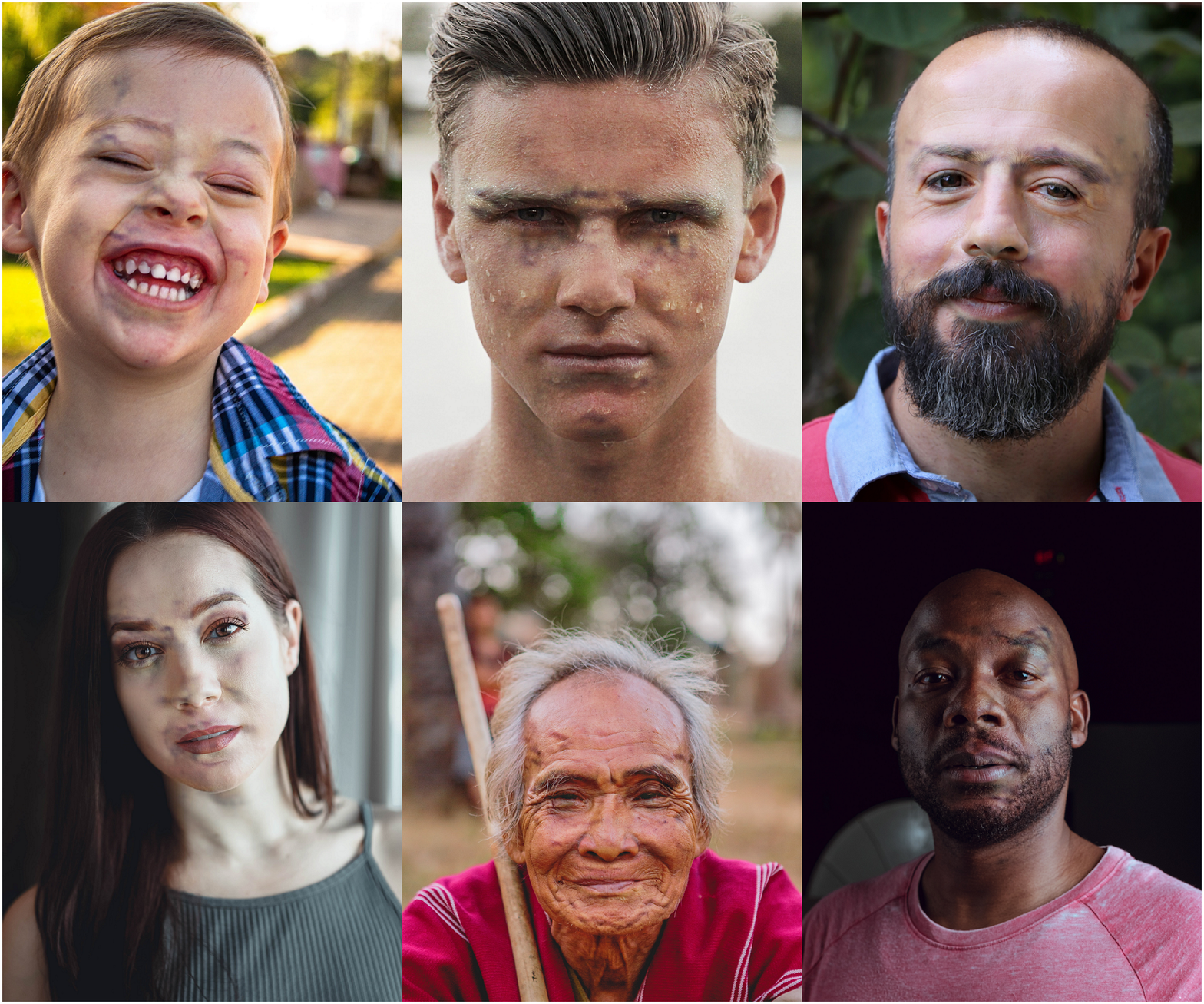}
			\end{minipage}
		}
		\centering
		\caption{Comparison between the cloaked and the uncloaked versions of high-resolution images. Note that there are wrinkles, shadows and irregular purple spots on faces of the cloaked images.}\label{TwoPhotos}
		
	\end{figure}

	A sample of the resulting images from this experiment is displayed in Figure~\ref{TwoPhotos}, these figures show images of the same users before (a) and after being cloaked by Fawkes (b). From these figures, we can easily observe significant differences with and without cloaking. Notably, there are many wrinkles, shadows and irregular purple spots on the boy's face in the cloaked image. This protection may result in the reluctance of users to post the cloaked images online.

\subsection{Countermeasures}
	Sybil accounts are fake or bogus identities created by a malicious user to inflate the resources and influence in a target community~\cite{yang2014uncovering}. A Sybil account, existing in the same online community, is a separate account to the original one of the user $U$, but the account, bolstering cloaking effectiveness, can be crafted to boost privacy protection in Fawkes when clean and uncloaked images are leaked for training~\cite{shan2020fawkes}. Fawkes modifies the Sybil images to protect the user's original images from being recognized. These Sybil images induce the model to be misclassified because they occupy the same area within the feature space of $U$'s uncloaked images. However, the feature space of cloaked images is vastly different from the originals. Sybil accounts are ineffective since the clean images are first cloaked before testing. Furthermore, these cloaked photos occupy a different area within feature space from the Sybil images as well as the clean images. To put it differently, no defense can be obviously offered irrespective of how many Sybil accounts the user can own, as cloaked images and uncloaked images occupy different feature spaces. 
	We are also able to increase the number of multi-cloaks $m$ in step with Fawkes to ensure the robustness of \textsc{Oriole} due to the white-box nature of the attack.

\section{Conclusion}
	In this work, we present \textsc{Oriole}, a novel system to combine the advantages of data poisoning attacks and evasion attacks to invalidate the privacy protection of Fawkes. To achieve our goals, we first train the face recognition model with multi-cloaked images and test the trained model with cloaked images. Our empirical results demonstrate the effectiveness of the proposed \textsc{Oriole} system. We have also identified multiple principle factors affecting the performance of the \textsc{Oriole} system. Moreover, we lay out the limitation of Fawkes and discuss it at length. We hope that the attack methodology developed in this paper will inform the security and privacy community of a pressing need to design better privacy-preserving deep neural models.

\section*{Acknowledgments}
The authors affiliated with East China Normal University were, in part, supported by 
NSFC-ISF Joint Scientific Research Program (61961146004) and Innovation Program of Shanghai Municipal Education Commission (2021-01-07-00-08-E00101). Minhui Xue was, in part, supported by the Australian Research Council (ARC) Discovery Project (DP210102670).

\bibliographystyle{splncs04}
\bibliography{name}
%
%
%
%


\end{document}